\documentclass[proof]{pasj00}
\draft

\begin{document}
\SetRunningHead{T. Sato et al.}{Suzaku X-ray Observation of 3C 391}
\Received{}
\Accepted{}

\title{Discovery of Recombining Plasma in \\the Supernova Remnant 3C 391}

\author{Tamotsu \textsc{SATO},\altaffilmark{1,2}
Katsuji \textsc{KOYAMA},\altaffilmark{3,4}
Tadayuki \textsc{TAKAHASHI},\altaffilmark{1,2}
Hirokazu \textsc{ODAKA},\altaffilmark{1}
and
Shinya \textsc{NAKASHIMA}\altaffilmark{1}}
\altaffiltext{1}{Department of High Energy Astrophysics, Institute of Space and Astronautical Science (ISAS), Japan Aerospace Exploration Agency (JAXA), 3-1-1Yoshinodai, Chuo, Sagamihara, Kanagawa, 252-5210, Japan}
\email{sato@astro.isas.jaxa.jp}
\altaffiltext{2}{Department of Physics, Graduate School of Science, University of Tokyo, Hongo 7-3-1, Bunkyo, Tokyo, 113-0033, Japan}
\altaffiltext{3}{Department of Earth and Space Science, Graduate School of Science, Osaka University, 1-1, Machikaneyama,Toyonaka, Osaka 560-0043, Japan}
\altaffiltext{4}{Department of Physics, Graduate School of Science, Kyoto University, Kita-Shirakawa, Sakyo-ku, Kyoto 606-8502, Japan}

\KeyWords{ISM: individual objects (3C 391, G31.9+0.0) --- ISM: supernova remnants --- X-rays: ISM}

\maketitle

\begin{abstract}
Recent X-ray study of middle-aged supernova remnants (SNRs) reveals strong radiative recombination continua (RRCs) associated with overionized plasmas, of which the origin still remains uncertain.
We report our discovery of an RRC in the middle-aged SNR 3C 391.
If the X-ray spectrum is fitted with a two-temperature plasma model in collisional ionization equilibrium (CIE), residuals of Si\emissiontype{XIV}~Ly$\alpha$ line at 2.006~keV, S\emissiontype{XVI}~Ly$\alpha$ line at 2.623~keV and the edge of RRC of Si\emissiontype{XIII} at 2.666~keV are found. 
The X-ray spectrum is better described by a composite model consisting of a CIE plasma and a recombining plasma (RP).
The abundance pattern suggests that the RP is associated to the ejecta from a core-collapse supernova with a 15\Mo~progenitor mass.
There is no significant difference of the recombining plasma parameters between the southeast region and the northwest region surrounded by dense molecular clouds.
We also find a hint of Fe\emissiontype{I}~K$\alpha$ line at 6.4~keV ($2.4\sigma$ detection) from the southeast region of the SNR.
\end{abstract}

\section{Introduction}
\label{sec:introduction}

The standard picture of thermal plasmas associated with supernova remnants (SNRs) is that the state of the gas appears to be either an ionizing plasma (IP) for young or intermediate aged SNRs (e.g.\ Kepler's SNR, Kinugasa and Tsunemi 1999), or plasma in collisional ionization equilibrium (CIE) for old SNRs (e.g. 30 Dor C, Yamaguchi et al. 2009). However, recent X-ray observations have revealed several SNRs are in recombination-dominated phase (IC 443, Yamaguchi \etal\ 2009; W49B, Ozawa \etal\ 2009; G359.1$-$0.5, Ohnishi \etal\ 2011; W28, Sawada \& Koyama 2012; W44, Uchida \etal\ 2012; G346.6$-$0.2, Yamauchi \etal\ 2013). These recombining plasmas (RPs) have an ionization temperature ($T_{\mathrm{z}}$) higher than the electron temperature ($T_{\mathrm{e}}$); $T_{\mathrm{z}}$ is a useful parameter defined to be the temperature which would be required to ionize the plasma to the same degree assuming that the plasma is in CIE.
The discovery of RPs at SNRs requires us to modify the conventional evolution scenario of SNRs.

3C 391 (G31.9+0.0) is a bright Galactic SNR displaying diffuse X-ray emission close the center of its radio shell and is classified as a mixed-morphology SNR (MM-SNR) (Reynolds \& Moffett 1993).
The radio shell with a radius of \timeform{5'} has the shape of a partial circle in the northwest (NW) part and there is a faint extended structure in the southeast (SE) region.
This suggests a break-out morphology into a lower density region from the main SNR shell.
The detection of two OH maser spots at 1720~MHz from 3C 391 by Frail \etal\ (1996) indicates an interaction with molecular clouds. They also reported a possibly extended OH emission at the edge of the NW shell.
The velocity of two maser spots (105 and 110~km~s$^{-1}$) agrees with a previous study of H\emissiontype{I} absorption velocity (Radhakrishnan \etal\ 1972).
H\emissiontype{I} observations show that the distance to 3C 391 is between 7.2~kpc and 11.4~kpc, assuming a distance to the Galactic center of 8.5~kpc.
The surrounding molecular gas of 3C 391 can be traced by CO (J $=$ 1$\rightarrow$0) emission at velocities from 90 to 110~km~s$^{-1}$ (Wilner \etal\ 1998). 
The results of the CO observation support the early expansion of 3C 391 inside a dense molecular cloud and the breakout of the blast wave to the SE region.

Knowledge of the astrophysical environment around SNRs is of great importance in order to reveal the origin of the RPs in the context of the SNR evolution.
Interestingly, most of the RP SNRs are MM-SNRs that interact with molecular clouds and are associated with bright GeV gamma-ray emissions (IC 443, Abdo \etal\ 2010a; W28, Abdo \etal\ 2010b; W49B, Abdo \etal\ 2010c; W44 Abdo \etal\ 2010d). While some young SNRs show bright TeV gamma-rays (RX J1713.7-3946, Aharonian \etal\ 2004), middle-aged SNRs are usually found to be predominantly emitting in the GeV energy band.
A significant excess (13$\sigma$) of GeV gamma-rays at the position of 3C 391 with Fermi-LAT was reported by Castro \& Slane (2010).
The radio rim of the NW region and the closer OH maser spot are \timeform{4'} apart from the GeV emission peak in the test statistic map.
The GeV emission associated with 3C 391 probably arises from accelerated cosmic rays in the shock-heated molecular clouds.
The interaction with the molecular gas and the association of the gamma-ray emission would be the keys for understanding the morphology and spectrum of the thermal plasmas in RP SNRs.

In this paper, we report on an X-ray study of SNR 3C 391, taking advantages of low and stable in-orbit background and high sensitivity of the Suzaku X-ray observatory (Mitsuda \etal\ 2007). We construct a model for the galactic background emission to complement the poor photon statistics, then we fit several plasma models to the X-ray spectrum of 3C 391. Our observation data is a part of the Suzaku Key Project to search for RPs in galactic SNRs. Section 2 describes the observation and data reductions and Section 3 presents the data analysis. Then, we interpret our results in Section 4 and conclude in Section 5.

\section{Observations and Data Reductions}
\label{sec:observation}

We observed SNR 3C 391 for an exposure time of 100 ks with the Suzaku satellite in 2010 October 22--24 during the AO5 cycle.
This observation was performed as one of the Key Project observations aimed at systematic search for RPs.
The pointing position was ($\alpha_{2000}$, $\delta_{2000}$) = (282.3812, $-$0.9417).
The observation instrument we used is the X-ray Imaging Spectrometer (XIS; Koyama \etal\ 2007), X-ray charge coupled device (CCD) cameras on the focal plane of the X-Ray Telescope (XRT; Serlemitsos \etal\ 2007). XIS\,0 and 3 are front-illuminated (FI) CCDs, and XIS\,1 is a back-illuminated (BI) CCD. 
The XIS was operated with the normal-clocking, full-frame mode.
Since in-orbit radiation damage has degraded the charge transfer efficiency of the XIS, the spaced-row charge injection technique is applied for the XIS.

We reprocessed the XIS data with the latest calibration results instead of using the cleaned events.
We used the HEADAS software of version 6.13 (Suzaku software of version 20). The calibration database (CALDB) released on 2013 March was used for data processing.
We adopted standard event selection criteria for Suzaku XIS data processing. We excluded events in the intervals of the South Atlantic Anomaly, low-Earth elevation angles less than 5 degrees and the day earth elevation angle less than 20 degrees.
After the data screening, the effective exposure time is about 99.4~ks. 
The data of the two editing mode, 3$\times$3 and 5$\times$5 pixel events, were combined, and then we removed flickering and hot pixels of the XIS.

\section{Analysis and Results}
\label{sec:analysis}
\subsection{X-ray Image}
\begin{figure}[htbp]
\begin{center}
\includegraphics[width=7cm]{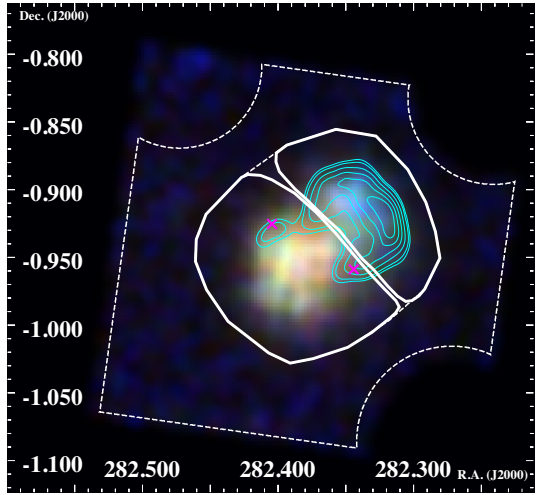}
\caption{
NXB-subtracted XIS\,3 image of 3C 391 after the correction of vignetting effect. The energy bands are 0.8--1.7~keV (red), 1.8--3.0~keV (green) and 3.0--5.0~keV (blue). The solid and dashed lines indicate the source regions (NW and SE) and background region, respectively. Contours show a flux map of 1.4~GHz by VLA (Condon et al. 1998). Locations of OH masers are plotted with the magenta crosses (Frail \etal\ 1996). Photons in 8 $\times$ 8 pixels (\timeform{8'.32} $\times$ \timeform{8'.32}) were summed and smoothed with a Gaussian kernel of $\sigma = \timeform{2.6'}$.
}
\label{fig:image XIS3}
\end{center}
\end{figure}

X-ray diffuse emission is clearly detected in the central region of the remnant and is encompassed by the radio rim in the NW.
Figure~\ref{fig:image XIS3} shows an X-ray image of 3C 391 (XIS\,3 only), after correcting the vignetting effect and subtracting the instrumental background (the so-called non X-ray background, or NXB). The NXB was estimated by using xisnxbgen (Tawa \etal\ 2008).

The northwest (NW) region associated with the radio rim has harder X-rays than the southeast (SE) region. 
It indicates the difference of the thermal plasma or circumstellar matters between those regions.
We thus divided the image into two regions: the NW and SE region.

\subsection{Background Estimation}

3C 391 is located on the Galactic plane, and hence the major background component is the Galactic ridge X-ray emission (GRXE), whose X-ray flux strongly depends on the position (e.g. Uchiyama \etal\ 2011, 2013). Since the photon statistics of the background region (surrounding region of 3C 391 in the same XIS field) is not good enough due to the limited exposure time and the effective area, subtracting the background data from the source data results in large statistical errors in the SNR spectrum. We therefore constructed a background model for the GRXE.  We extracted the background spectrum from the region excluding 3C 391 in the same field of view, and fitted the GRXE model compiled by Uchiyama \etal\ (2013) to the spectrum. 
In this background region, a few percent level of photon contamination from 3C 391, which is due to the leakage of the XRT point spread function into the background region, cannot be neglected. 

\subsubsection{Phenomenological Model Fit}
In order to construct the background model, we need to make phenomenological models both for the source and the background regions.
We extracted spectra from the whole 3C 391 region and the background region after subtracting the NXB.
After the vignetting correction, we subtracted the background spectrum from the 3C 391 spectrum. 
The redistribution matrix file (RMF) and the ancillary response file (ARF) were created by xisrmfgen and xissimarfgen, respectively (Ishisaki \etal\ 2007).
We used the spectral analysis software XSPEC version 12.8.1 (Arnaud 1996).
We simultaneously fitted all the XIS\,0, 1, and 3 data with a single model. The energy band from 1.7 to 1.8~keV were ignored because of the calibration uncertainty around the Si K edge.

\begin{figure}[htbp]
\begin{center}
\includegraphics[width=7cm]{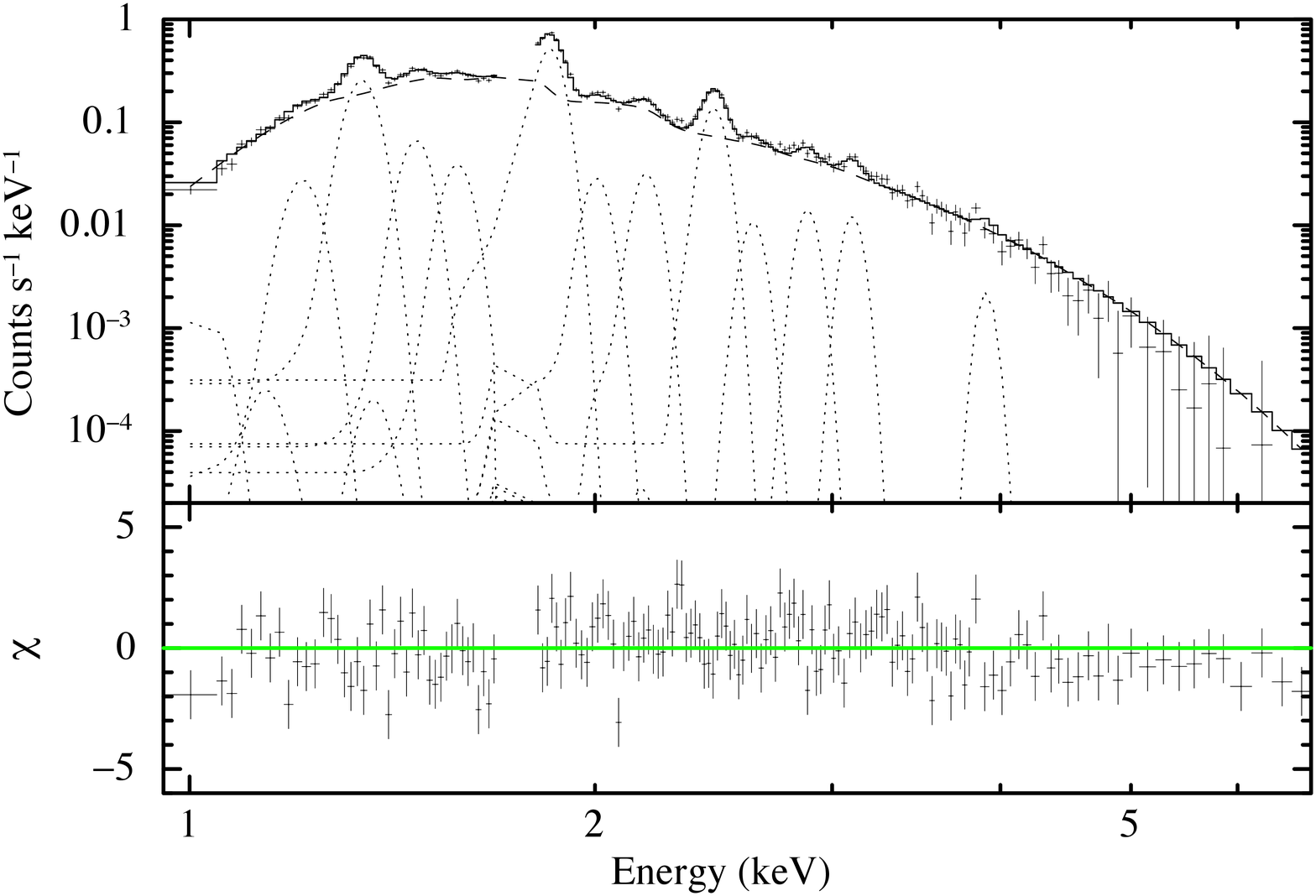}
\caption{X-ray spectrum of 3C 391 with the best-fit phenomenological model (solid line). Only a merged spectrum of 
FI CCDs are shown for visibility. The dash line shows bremsstrahlung emission and the dotted lines correspond to atomic line emissions.}
\label{fig:spec phenom}
\end{center}
\end{figure}

We then modeled the spectra by using a phenomenological model: the continuum X-ray emission is described by bremsstrahlung with temperature of $kT_{\mathrm{b}}$ and line emissions from various ionized atoms are given by Gaussian lines. The interstellar photoelectric absorption was estimated using the cross sections by Morrison and McCammon (1983) with H\emissiontype{I} column density $N_{\mathrm{H}}$. 
This phenomenological model successfully reproduces the data ($\chi^2 =$ 404.01/307). Figure~\ref{fig:spec phenom} and table~\ref{tbl:par phenom} show the best-fit spectrum and parameters, respectively. 

\begin{table}[htbp]
\caption{Parameters of the phenomenological model.$^*$}
\label{tbl:par phenom}
\begin{center}
\begin{tabular}{lccc}
\hline

Parameter & \multicolumn{2}{c}{bremsstrahlung} \\ \hline
$N_{\mathrm{H}}$ (10$^{22}$~cm$^{-2}$) & \multicolumn{2}{c}{1.90$\pm$0.05} \\
$kT_{\mathrm{b}}$ (keV) & \multicolumn{2}{c}{0.671$\pm$0.014} \\ \hline
Emission line & energy (keV)$^\dag$ & intensity$^\ddag$  \\ \hline
Ne~\emissiontype{X}~Ly$\alpha$ & 1.022 & 11$\pm$10\\
Ne~\emissiontype{X}~Ly$\beta$ & 1.211 & 23$\pm$8 \\
Mg~\emissiontype{XI}~K$\alpha$ & 1.341$\pm$0.001 & 103$\pm$9 \\
Mg~\emissiontype{XII}~Ly$\alpha$ & 1.472 & 14.7$\pm$2.6 \\
Mg~\emissiontype{XI}~K$\beta$ & 1.579 & 7.1$\pm$1.9 \\
Si~\emissiontype{XIII}~K$\alpha$ & 1.850$\pm$0.002 & 90$\pm$3 \\
Si~\emissiontype{XIV}~Ly$\alpha$ & 2.006 & 3.6$\pm$0.8 \\
Si~\emissiontype{XIII}~K$\beta$ & 2.183 & 3.2$\pm$0.5 \\
S~\emissiontype{XV}~K$\alpha$ & 2.451$\pm$0.002 & 15.9$\pm$0.6 \\
S~\emissiontype{XVI}~Ly$\alpha$ & 2.621 & 1.1$\pm$0.4 \\
S~\emissiontype{XV}~K$\beta$ & 2.877 & 1.3$\pm$0.3 \\
Ar~\emissiontype{XVII}~K$\alpha$ & 3.107 & 1.02$\pm$0.24 \\
Ca~\emissiontype{XIX}~K$\alpha$ & 3.903 & 0.16$\pm$0.13 \\ \hline
$\chi^2$/d.o.f & \multicolumn{2}{c}{404.01/307} \\ \hline

\end{tabular}
\end{center}
$^*$ The uncertainties are the 90\% confidence range. \\
$^\dag$ The centroid energies of strong K$\alpha$ lines are free because they are fundamentally shifted by the contribution of another faint emission lines. \\
$^\ddag$ The units are 10$^{-5}$~photons~cm$^{-2}$~s$^{-1}$.
\end{table}

\subsubsection{Construction of the Background Model}
\label{sec:bgdmodel}

\begin{figure}[htbp]
\begin{center}
\includegraphics[width=7cm]{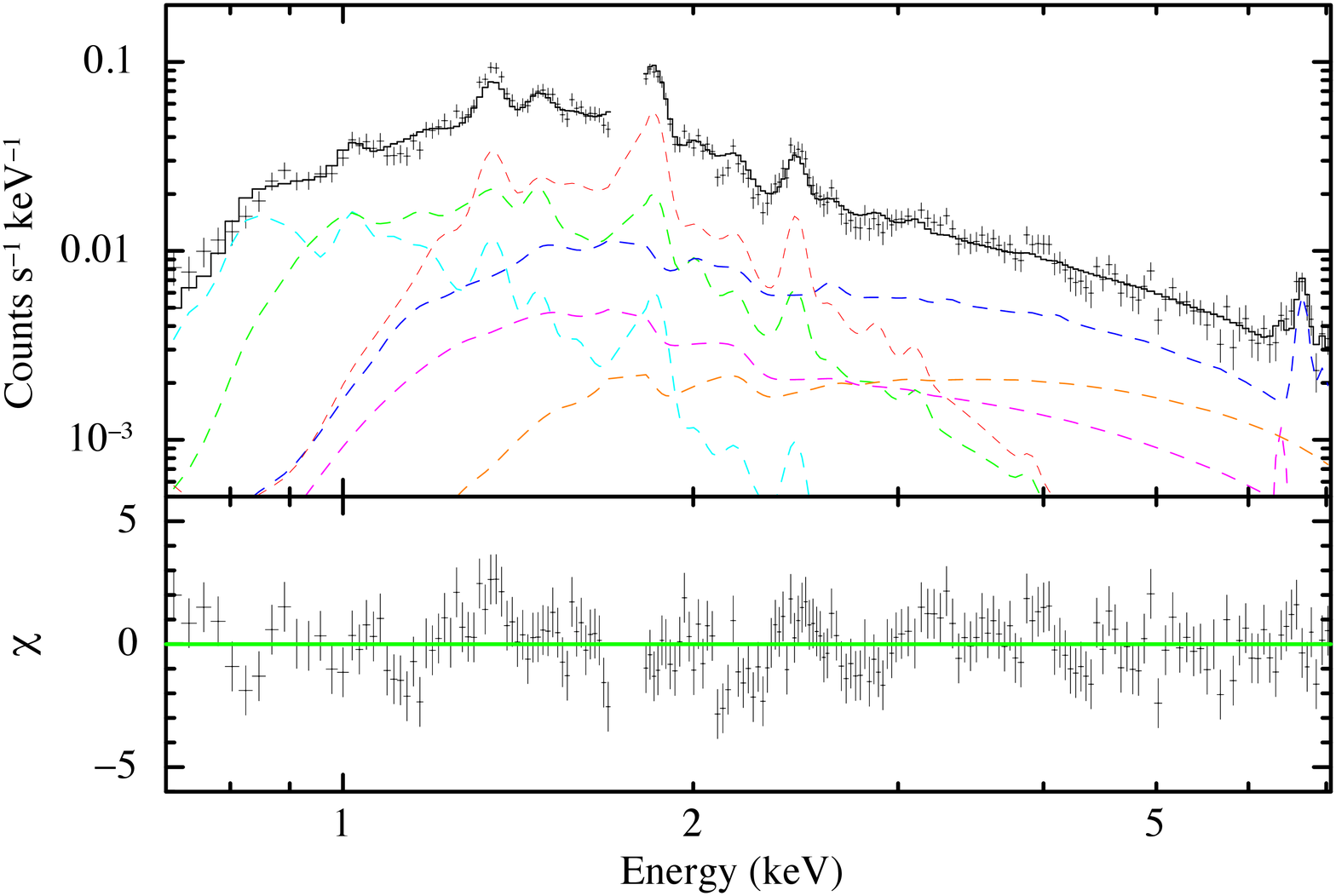}
\caption{NXB-subtracted spectrum extracted from the background region (black crosses). The best-fit model is shown with the solid line. For visibility, only merged FI data are displayed. The dash lines show the background components of FG (cyan), LP (green), and HP (blue). The magenta and orange lines shows Fe\emissiontype{I}~K$\alpha$ and CXB, respectively. The contamination spectrum from 3C 391 is shown by the red line.}
\label{fig:spec background}
\end{center}
\end{figure}

To make the GRXE model in the 3C 391 region, we fitted a composite model consisting of the GRXE, the cosmic X-ray background (CXB) and the contamination of 3C 391 spectrum to the NXB-subtracted background spectrum.
For the CXB, we used a power-law function with a photon index 1.4, which has a flux of 8.2 $\times$ 10$^{-7}$~photons~cm$^{-2}$~s$^{-1}$~arcmin$^{-2}$~keV$^{-1}$ at 1~keV (Kushino \etal\ 2002).  For the contamination spectrum, we used the phenomenological model given in the previous section (table 1) with the normalization as a free parameter. 
Uchiyama \etal\ (2013) composed the GRXE with a foreground plasma (FG), a low temperature plasma (LP) and a high temperature plasma (HP). All these models are thin thermal plasmas in CIE, whose temperatures are designated as $kT_{\mathrm FG}$, $kT_{\mathrm LP}$ and $kT_{\mathrm HP}$, respectively.  We used the APEC model by Smith \etal\ (2001).  Metal abundances were taken relative to the solar values (Anders and Grevesse 1989). The GRXE also includes a cold matter (CM) emission with neutral iron line (6.4~keV) of equivalent width $\sim450$~eV. Thus the GRXE model is given by;
\begin{equation}
A_{\mathrm{FG}} \times {\mathrm{FG}} + A_{\mathrm{GRXE}} \times ({\mathrm{LP + HP + CM}}) + A_{\mathrm{CXB}} \times {\mathrm{CXB}}, 
\end{equation}
where $A_{\mathrm{FG}}$ and $A_{\mathrm{GRXE}}$ are absorptions by the Galactic H\emissiontype{I} with column densities $N_{\mathrm{H}}$ for the foreground plasma and GRXE components, respectively. An interstellar absorption of CXB ($A_{\mathrm{CXB}}$) was taken to be twice of $A_{\mathrm{GRXE}}$.  Following  the result of Uchiyama \etal\ (2013), $A_{\mathrm{FG}}$ and $kT_{\mathrm{FG}}$ were fixed to be $0.56\times10^{22}$~cm$^{-2}$ and 0.59~keV, respectively.  Although the foreground plasma has another CIE component with a temperature of $kT \sim 0.09$~keV, we neglected this component because no significant contribution was expected in our energy range. 
Figure~\ref{fig:spec background} shows the background spectrum with the best-fit background model including contamination spectrum. The best-fit parameters of the GRXE are given in table~\ref{tbl:par background}. 

\begin{table}[htbp]
\caption{Fitting parameters of the background region$^*$}
\begin{center}
\begin{tabular}{ccc}
\hline

Parameter & LP & HP \\ \hline

$N_{\mathrm{H}}$ (10$^{22}$~cm$^{-2}$) & \multicolumn{2}{c}{ 1.13 $\pm$ 0.10 } \\
$kT$ (keV) & $0.87^{+0.05}_{-0.06}$ & $6.7^{+1.5}_{-1.2}$ \\
Abundance (solar) & $0.35^{+0.22}_{-0.13}$ & $0.78^{+0.36}_{-0.21}$ \\ \hline
$f_{6.4}$$^\dag$ & \multicolumn{2}{c}{ 1.5 $\pm$ 1.1 } \\ \hline 
$\chi^2$ / d.o.f & \multicolumn{2}{c}{ 458.73/312 } \\ \hline

\end{tabular}
\end{center}
$^*$ The uncertainties are the 90\% confidence range. \\
$^\dag$Observed photon flux of 6.4~keV in units of 10$^5$~photons~cm$^{-2}$~s$^{-1}$.
\label{tbl:par background}
\end{table}

\subsection{Model Fits to the Whole 3C 391 Spectrum}

We fitted several plasma models to the NXB-subtracted spectrum in the entire region of 3C 391. The GRXE model function determined in the previous section was included in the model fitting.
The abundances of Ne, Mg, Al, Si, S, Ar, Ca and Fe were free parameters. 
We first tried the APEC model in the XSPEC package, a single-temperature ($kT$) optically thin thermal plasma model in CIE state. 
This model is rejected with a large $\chi^2$ / d.o.f value (565.83/315). 

One of the most prominent residuals is the excess below  $\sim$1.5~keV. We therefore added another APEC model with metal abundances treated as free parameters. The best-fit metal abundances were all consistent with 1 solar.  We hence fixed all the  abundances at 1 solar.
Then the fit is improved  to $\chi^2$ / d.o.f $=$ 397.51/313 (1.27).

In the atomic database of the current plasma codes, the higher Rydberg series of Fe-L lines ($\sim$1.2~keV) are incomplete (Brickhouse \etal\ 2000). If we add a $\sim$1.2 keV Gaussian line, $\chi^2$ / d.o.f is well improved from $\sim$1.27 to $\sim$1.25. The best-fit spectrum and parameters are given in figure 4 and table 3 (left column).

This two-temperature CIE model still leaves significant residuals at Si\emissiontype{XIV}~Ly$\alpha$ line (2.006~keV), hump-like feature at $\sim2.5$~keV and a hint of excess at 6.4 keV.  The data excess at the 2.5 keV bump above the applied model is $2.6\pm0.7 \times 10^{-3}$~counts~s$^{-1}$ (2.4--2.8~keV).  In the background spectrum, the data excess in the 2.4--2.8~keV band is negative and as small as $0.32\pm0.19 \times 10^{-3}$~counts~s$^{-1}$. 
Thus the excess at 2.5 keV in the SNR spectrum, after the background subtraction, is not an artifact due to the background subtraction, but is a real structure. Since the bump energy of $\sim2.5$~keV corresponds to S\emissiontype{XVI}~Ly$\alpha$ line (2.623~keV) and the edge of radiative recombination continuum (RRC) of He-like Si (2.666 keV; Yamaguchi \etal\ 2009), this is likely due to a recombining plasma (e.g. Sawada and Koyama 2012). 
We thus applied an NEIJ model in the SPEX version 2.03.03 (Kaastra \etal\ 1996) for the high temperature component.
This model describes a transition plasma that initially in the CIE state with the temperature $kT_1$ then only the electron temperature dropped to $kT$.  The recombination phase is given by the parameter $n_{\mathrm{e}}t$, where $n_{\mathrm{e}}$ and $t$ are the number density of electrons and the elapsed time after the electron cooling, respectively. 

\begin{figure*}[htbp]
\begin{center}
\includegraphics[width=7cm]{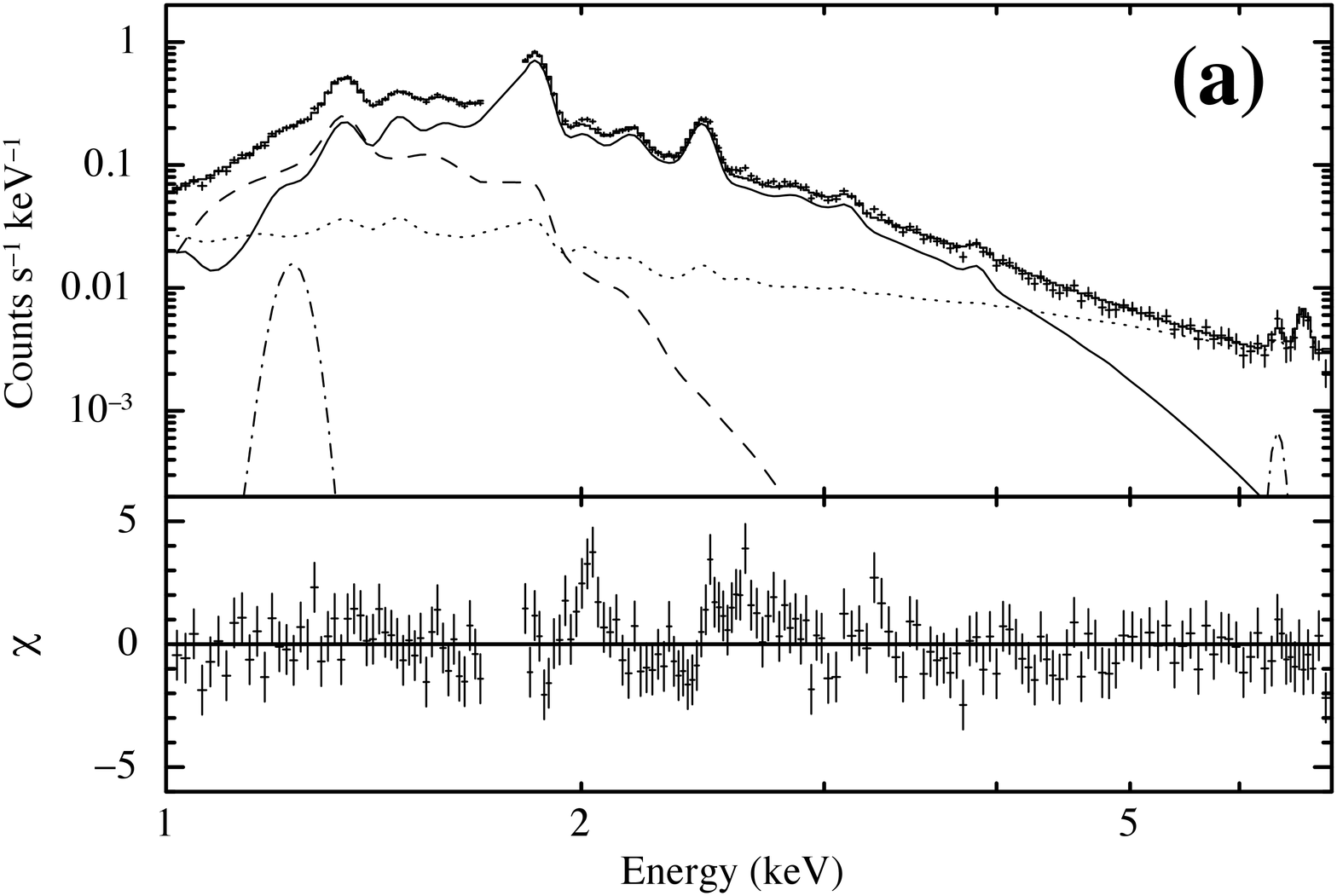}
\includegraphics[width=7cm]{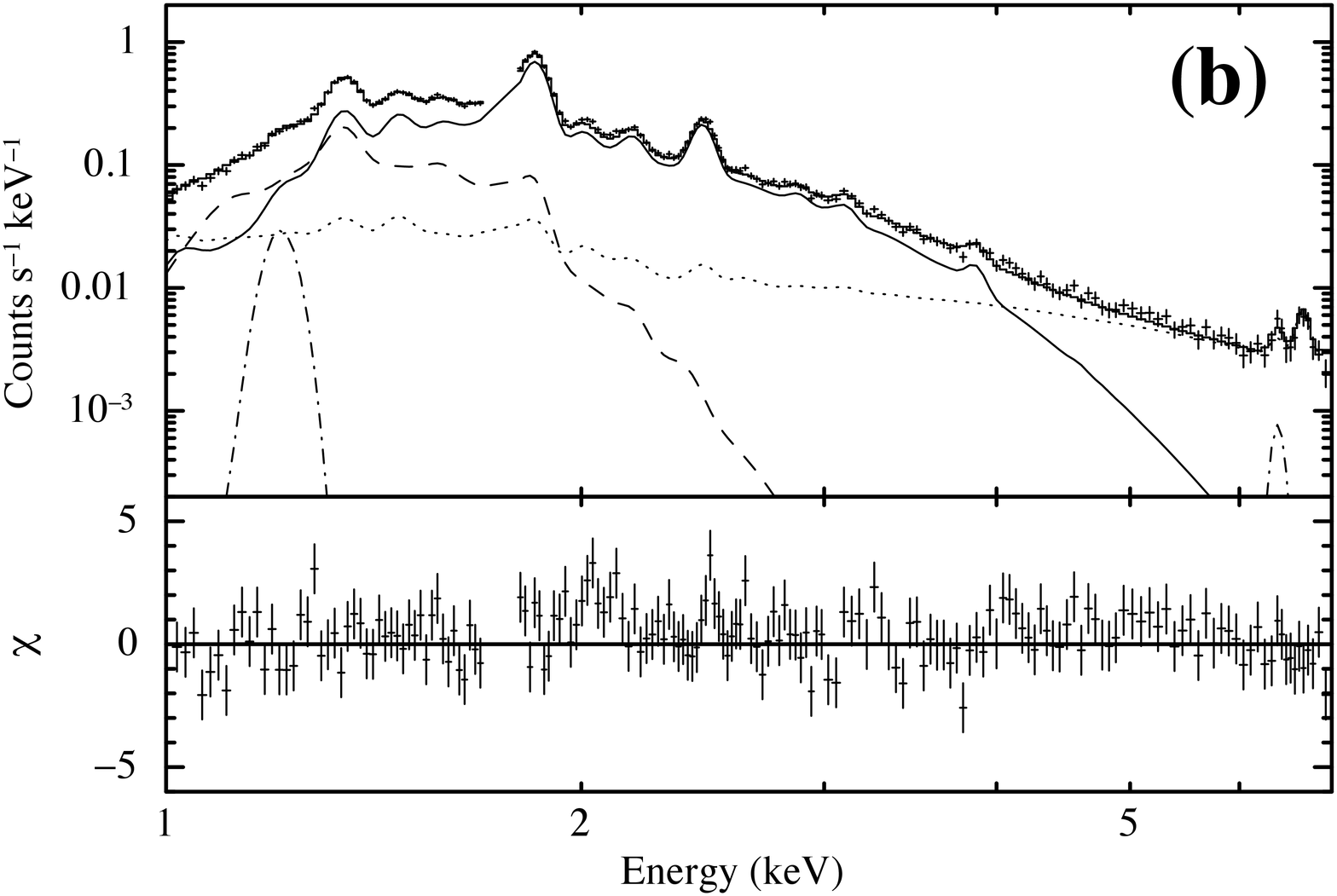}
\caption{NXB subtracted spectra of 3C 391 with the best-fit models and the residuals (lower panel) of the APEC+APEC model (a) and CIE+NEIJ model (b). For visibility, only merged FI spectrum is displayed. The solid and dashed lines correspond to the high and low temperature plasma models. The background model and the Gaussian functions are shown with the dashed and dash-dotted lines, respectively.}
\label{fig:spec}
\end{center}
\end{figure*}

The fit is largely improved from the two-temperature CIE model.  As is shown in figure~\ref{fig:spec}, the Si\emissiontype{XIV}~Ly$\alpha$ line (2.006~keV) and hump-like feature at $\sim2.5$~keV have disappeared, even though a hint of excess at 6.4 keV is still remained. 
The $\chi^2$ / d.o.f becomes almost acceptable level of 355.14/315 $\sim$ 1.13. The best-fit parameters are summarized in table~\ref{tbl:par all} (right column).

\begin{table}[htbp]
\caption{Best-fit parameters of the whole region$^*$}
\begin{center}
\begin{tabular}{ccccc}
\hline

 & \multicolumn{2}{c}{CIE Plasma}  & \multicolumn{2}{c}{Recombining Plasma} \\ \hline
Parameter & CIE & CIE & CIE & NEIJ \\ \hline
$N_{\mathrm{H}}$ (10$^{22}$~cm$^{-2}$) &  \multicolumn{2}{c}{2.88$\pm$0.08} &  \multicolumn{2}{c}{3.13$^{+0.14}_{-0.12}$} \\

$VEM$$^\dag$ (10$^{60}$~cm$^{-3}$) & 3.2$^{+2.3}_{-0.8}$ & 0.092$^{+0.011}_{-0.015}$ & 2.8$^{+1.4}_{-0.9}$ & 0.18$^{+0.03}_{-0.05}$ \\
$kT_1$ (keV) & -- & -- & -- & 1.8$^{+1.6}_{-0.6}$ \\
$kT$ (keV) & 0.172$^{+0.006}_{-0.017}$ & 0.597$^{+0.011}_{-0.008}$ & 0.170$^{+0.009}_{-0.008}$ & 0.495$\pm$0.015  \\
$n_{\mathrm{e}} t$ (10$^{11}$~cm$^{-3}$~s) & -- & -- & -- & 14.0$^{+1.5}_{-2.2}$  \\ \hline
Abundance (solar) \\ \hline
Ne & & 1.9$^{+1.0}_{-0.7}$ & & 0.7$^{+1.1}_{-0.4}$ \\
Mg & & 0.80$^{+0.18}_{-0.13}$ & & 0.66$^{+0.30}_{-0.12}$ \\
Al & & 0.8$\pm$0.3 & & 0.4$\pm$0.3 \\
Si & & 0.96$^{+0.17}_{-0.11}$ & & 0.82$^{+0.30}_{-0.12}$ \\
S & 1(fix) & 0.85$^{+0.16}_{-0.10}$ & 1(fix) & 0.81$^{+0.31}_{-0.11}$ \\
Ar & & 0.61$^{+0.17}_{-0.14}$ & & 0.65$^{+0.25}_{-0.16}$ \\
Ca & & 1.0$\pm$0.5 & & 1.6$^{+0.8}_{-0.5}$ \\
Fe & & $<$0.13 &  & $<$0.05 \\
others & & 1(fix) & & 1(fix) \\ \hline
$\chi^2$ / d.o.f & \multicolumn{2}{c}{389.68/311} & \multicolumn{2}{c}{355.14/315} \\ \hline
\end{tabular}
\end{center}
$^*$ The uncertainties are the 90\% confidence range. \\
$^\dag$ Volume emission measure $\int n_{\mathrm{e}} n_{\mathrm{p}} \mathrm{d}V$ at the distance of 8~kpc, where $n_{\mathrm{e}}$, $n_{\mathrm{p}}$ and $V$ are the electron and proton number densities (cm$^{-3}$) and the emitting volume (cm$^{3}$), respectively.\\
\label{tbl:par all}
\end{table}

Since the Chandra observation shows that the spectra of 3C 391 has spatial variations with many NEI plasmas (Chen \etal\ 2004), we examine the excess at around 2.5~keV in the context of a composite plasma model. We made a composite model including all NEI plasmas resolved by Chen \etal\ 2004. Their emission measures were fixed to their individual values. The ionization timescales determined by Chen \etal\ 2004 were fixed, and other timescales were free. $N_{\mathrm{H}}$ and abundances were free parameters. We also added the CIE plasma model with a low temperature ($<$0.2~keV).
This composite model gives the best-fit reduced $\chi^2$ of 1.47, leaving the excess of $\sim$2.5~keV of $2.1\pm0.7 \times 10^{-3}$~counts~s$^{-1}$ (2.4--2.8~keV). 
Also the excess of 6.4 keV line does not disappear.
If we replace the major 2-NEI plasmas with high emission measures (each $\sim$20\% of the total) with a 1-RP model, the reduced $\chi^2$ value improves to 1.15.  Since the best-fit reduced $\chi^2$ for the multi-NEI+CIE+RP is not smaller than that of the CIE+RP model, we use the CIE+RP model as the best approximation for the SNR spectra, and proceed with the following analysis and discussion.

\subsection{Spatial Distribution of the Recombining Plasma}

In figure~\ref{fig:image XIS3}, we can see significant difference between the NW and SE regions.  We therefore analyzed the NW and SE spectra separately using the same model as was applied to the spectrum of the entire region of 3C 391.

\begin{figure*}[htbp]
\begin{center}
\includegraphics[width=7cm]{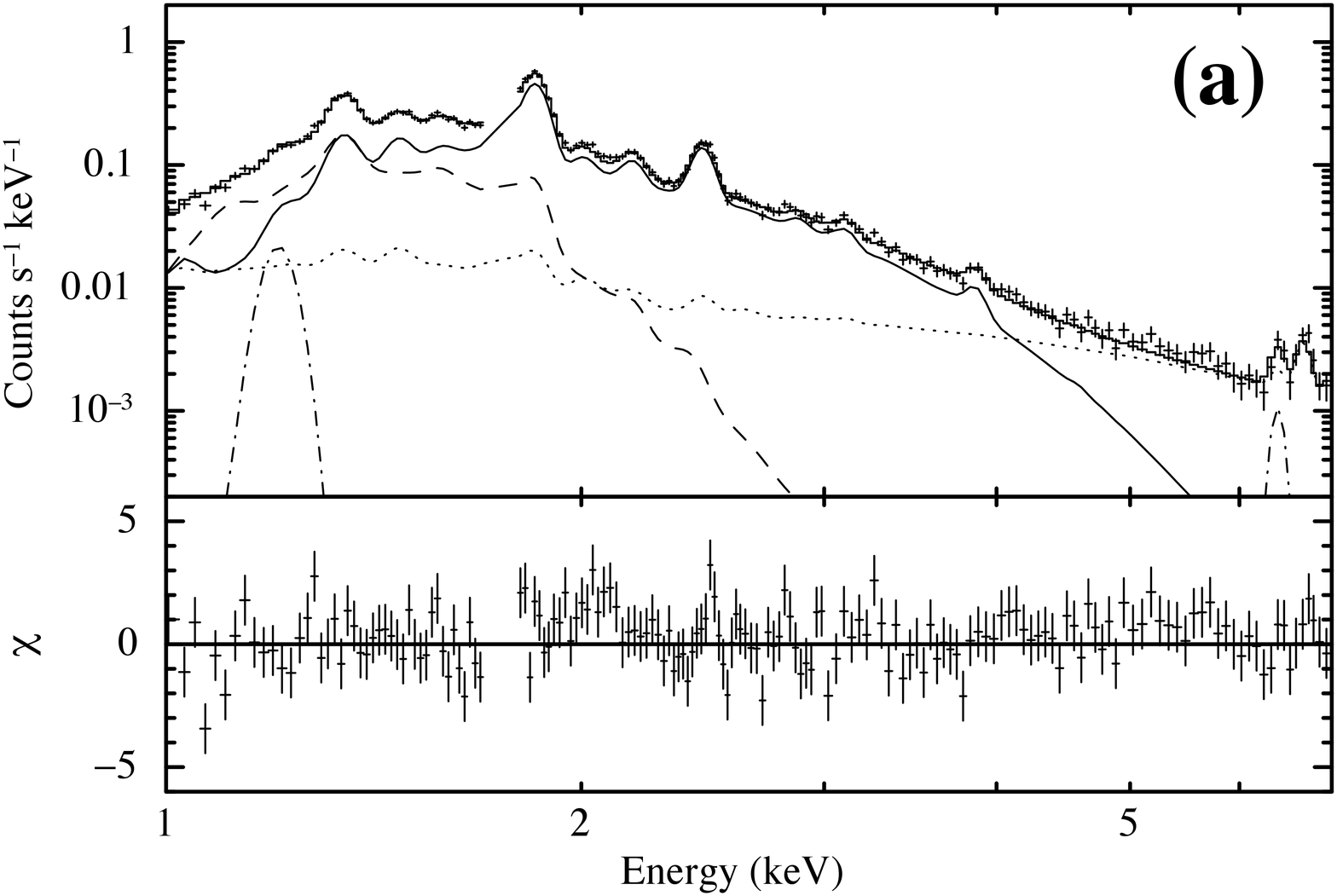}
\includegraphics[width=7cm]{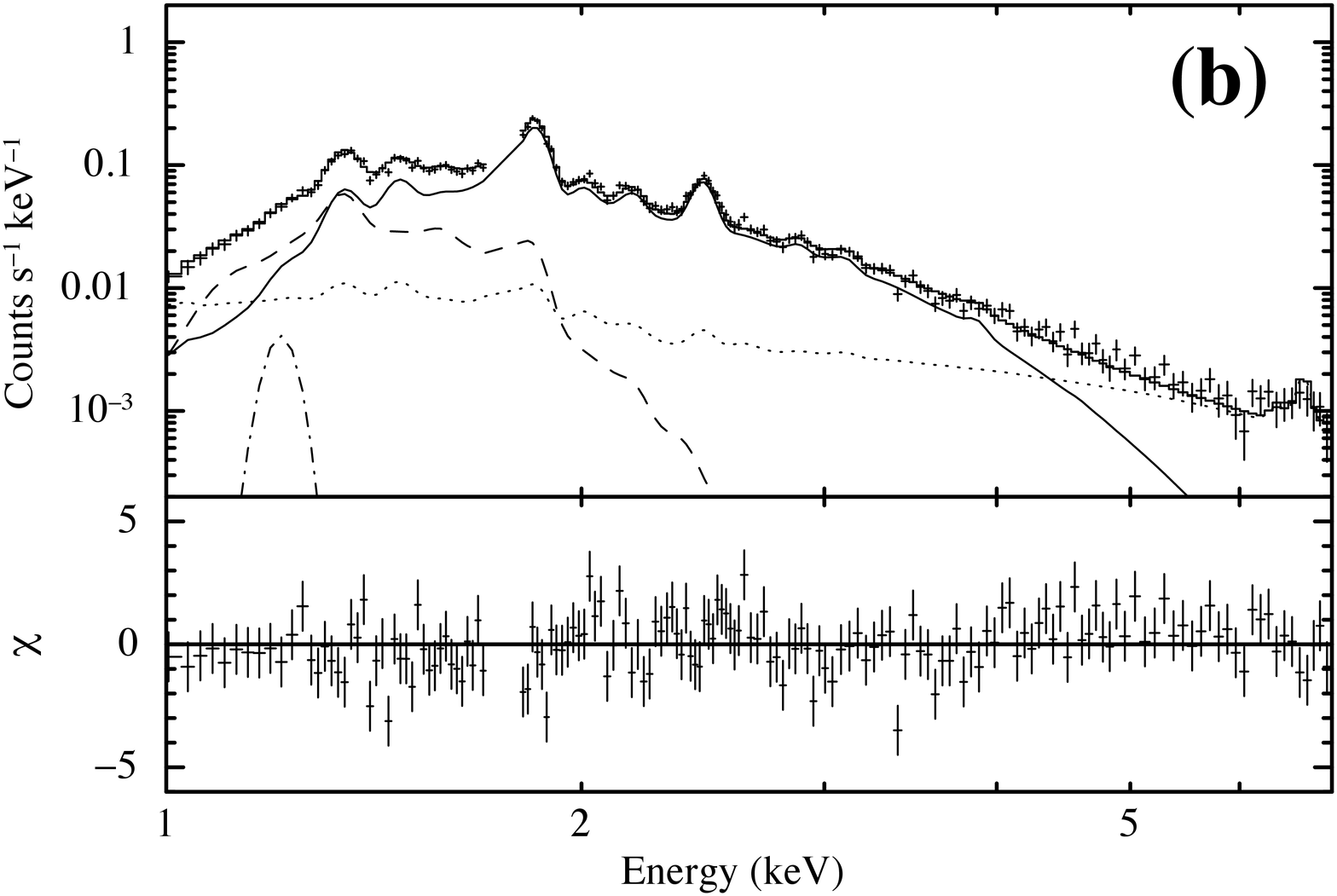}
\caption{NXB subtracted spectra of the SE region (a) and the NW region (b) with the best-fit models and these residuals. 
For visibility, only merged FI spectrum is displayed. Solid and dashed lines correspond with NEIJ and CIE model.
The background model and Gaussian functions are shown with the dashed, doted dashed-dotted lines, respectively.}
\label{fig:spec dist}
\end{center}
\end{figure*}

The fit for the NW and SE regions are marginally acceptable with $\chi^2/d.o.f$ of 333/275=1.21, and 306/266=1.15, respectively. However, we find no significant spectral differences except the possible presence of the 6.4 keV line in NW of which the significance level is $\sim 2.4 \sigma$ and a slightly larger $N_{\mathrm{H}}$ in SE.
Figure~\ref{fig:spec dist} (a) and (b) show the results of the NW and SE regions, respectively. The best-fit parameters of each region are shown in table~\ref{tbl:par dist}.

\begin{table}[htbp]
\caption{Best-fit parameters of each regions$^*$}
\begin{center}
\begin{tabular}{ccccc}
\hline

& \multicolumn{2}{c}{SE}  & \multicolumn{2}{c}{NW} \\ \hline
Parameter & CIE & NEIJ & CIE & NEIJ \\ \hline
$N_{\mathrm{H}}$ (10$^{22}$~cm$^{-2}$) &  \multicolumn{2}{c}{3.05$^{+0.12}_{-0.14}$} &  \multicolumn{2}{c}{3.45$^{+0.15}_{-0.16}$} \\

$VEM$$^\dag$ (10$^{60}$~cm$^{-3}$) & 1.5$^{+0.7}_{-0.5}$ & 0.09$^{+0.02}_{-0.03}$ & 2.3$^{+0.4}_{-1.0}$ & 0.08$\pm$0.02 \\
$kT_1$ (keV) & -- & 1.6$^{+1.9}_{-0.5}$ & -- & 2.7$^{+\infty}_{-2.0}$ \\
$kT$ (keV) & 0.184$^{+0.012}_{-0.010}$ & 0.50$\pm$0.02 & 0.159$^{+0.014}_{-0.005}$ & 0.55$\pm$0.02  \\
$n_{\mathrm{e}} t$ (10$^{11}$~cm$^{-3}$~s) & -- & 14$^{+2}_{-3}$ & -- & 15$\pm$2  \\ \hline
Abundance (solar) \\ \hline
Ne & & 1.3$^{+1.9}_{-0.8}$ & & 0.5$^{+1.3}_{-0.4}$ \\
Mg & & 0.85$^{+0.51}_{-0.22}$ & & 0.56$^{+0.31}_{-0.17}$ \\
Al & & 0.6$^{+0.4}_{-0.5}$ & & 0.2$^{+0.4}_{-0.1}$ \\
Si & & 1.1$^{+0.6}_{-0.2}$ & & 0.62$^{+0.24}_{-0.10}$ \\
S & 1(fix) & 1.1$^{+0.6}_{-0.2}$ & 1(fix) & 0.59$^{+0.24}_{-0.12}$ \\
Ar & & 0.9$^{+0.5}_{-0.3}$ & & 0.32$^{+0.23}_{-0.18}$ \\
Ca & & 2.2$^{+1.3}_{-0.8}$ & & 0.5$^{+0.6}_{-0.4}$ \\
Fe & & $<$0.08 &  & $<$0.5 \\
others & & 1(fix) & & 1(fix) \\ \hline
$f_{6.4}$$^\ddag$ & \multicolumn{2}{c}{2.1$\pm$0.9} & \multicolumn{2}{c}{$<$0.5} \\ \hline
$\chi^2$ / d.o.f & \multicolumn{2}{c}{332.72/275} & \multicolumn{2}{c}{306.32/266} \\ \hline

\end{tabular}
\end{center}
$^*$ The uncertainties are the 90\% confidence range. \\
$^\dag$ Volume emission measure $\int n_{\mathrm{e}} n_{\mathrm{p}} \mathrm{d}V$ at the distance of 8~kpc, where $n_{\mathrm{e}}$, $n_{\mathrm{p}}$ and $V$ are the electron and proton number densities (cm$^{-3}$) and the emitting volume (cm$^{3}$), respectively.\\
$^\ddag$ The units are 10$^{-8}$~photons~cm$^{-2}$~s$^{-1}$~arcmin$^{-2}$. The uncertainties are the 1 $\sigma$ confidence range.  \\
\label{tbl:par dist}
\end{table}

\section{Discussion}
\label{sec:discussion}

\begin{figure}[htbp]
\begin{center}
\includegraphics[width=9cm]{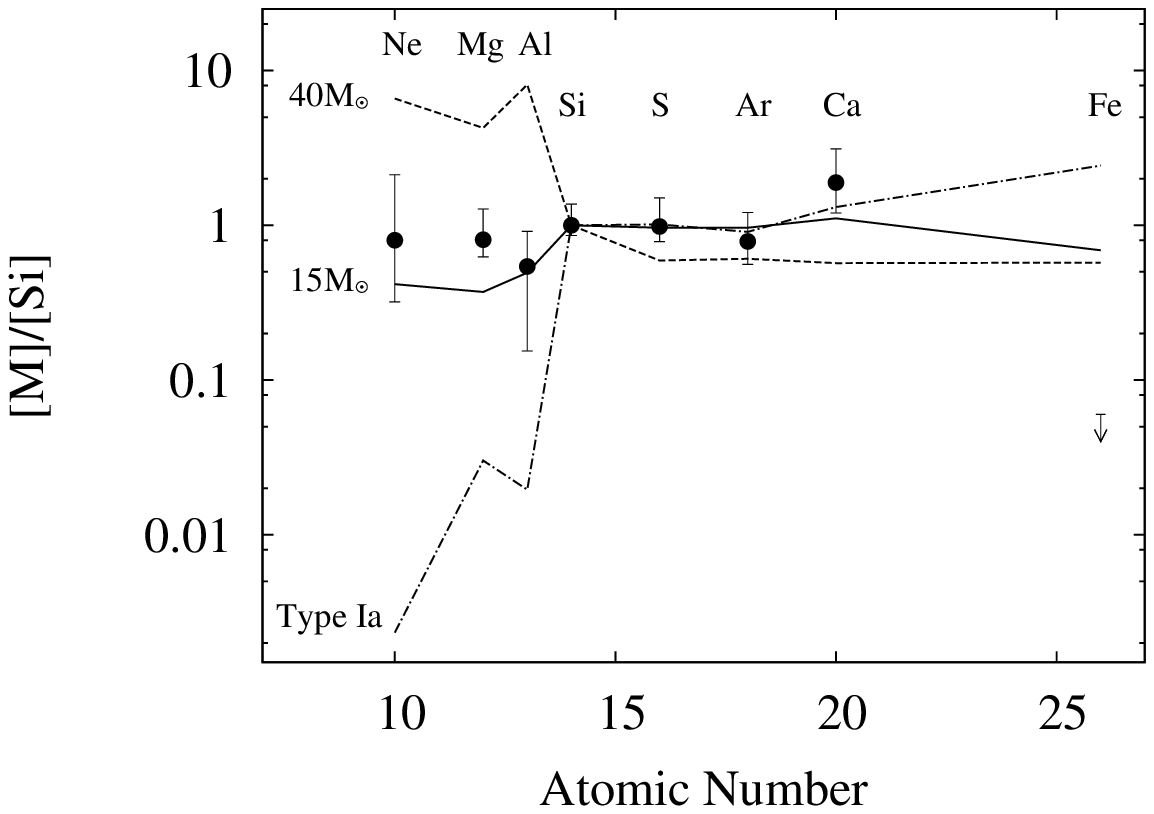}
\caption{Best-fit metal abundances of 3C 391 relative to Si abundance as a function of atomic number. The dashed-dotted line shows Type Ia SN model (CDD1; Iwamoto et al. 1999). The solid and dashed lines represent CC SN models with progenitor masses of 15~\Mo and 40~\Mo, respectively (Woosley \& Weaver 1995).}
\label{fig:mass}
\end{center}
\end{figure}

The X-ray spectrum of 3C 391 is represented by two plasma models: a low temperature CIE and a high temperature RP. The CIE plasma is probably emitted by the ISM heated by the blast waves, because it has 1 solar abundances. The abundances of the  RP are given in figure~\ref{fig:mass}, together with some theoretical model of Ia (Iwamoto et al. 1999) and core-collapse supernova (CC SN; Woosely and Weaver 1995). Although the error is large, the abundance pattern except Fe is similar to a CC SN of progenitor mass of $\sim$15~\Mo.
Thus we regard that the RP has an ejecta origin from a CC SN of a $\sim$15~\Mo~progenitor star.
In Section 4.1, we discuss consistency of the observed data with this assumption. 

The abundance of Fe in ejecta largely depends on the mass cut of the CC SN.
If a major fraction of Fe would be captured into a collapsed object (neutron star or black hole) in the case of 3C 391, a small abundance of Fe in the ejecta would be reasonable, unless the explosion is highly asymmetric as is  demonstrated by Yasumi et al. 2014. 
Another possibility is that only a small fraction of the Fe-rich ejecta has been shocked by the reverse shock (Fe-rich ejecta are expected to be located at the very inner part of the remnant).
Since the Fe abundance estimated using the Fe-L lines only has a large uncertainty due to their incompleteness, we can not conclude why the abundance of Fe is lower than expected.

The recombination parameter ($n_{\mathrm{e}}t \sim 1.4 \times 10^{12}$~cm$^{-3}$~s) is the largest among any RP discovered so far in MM-SNRs 
(Yamaguchi \etal\ 2009, 2012, Ozawa \etal\ 2009, Ohnishi \etal\ 2011, Sawada \& Koyama 2012, Uchida \etal\ 2012, Yamauchi \etal\ 2013).
The spectra of the SE and NW also show the RP. Since the extended OH emission reported in the NW region suggests the interaction with molecular clouds, the slightly higher $N_{\mathrm{H}}$ than the SE would be reasonable. 
On the other hand, $n_{\mathrm{e}}t$ and $kT$ show no significant difference between the two regions. 

We found a hint of line emission at 6.4 keV, Fe\emissiontype{I}~K$\alpha$  line at $2.4\sigma$ level from the SE region, and not from the NW region where the SNR interacts with molecular clouds. 
However the SE region is coincident with OH masers. One possibility is that the  Fe\emissiontype{I}~K$\alpha$ 
line is due to low energy cosmic-ray electrons, which are preferentially made in the SE region.
The gyro radius of the lower energy electrons is extremely small, and hence diffusion 
length is far smaller than the SNR size. The absence of the Fe \emissiontype{I}~K$\alpha$ line at the NW, where cool gas is denser than the SE, would be due to this limited diffusion length.
Deeper observations of hard X-ray will probably detect bremsstrahlung by low energy cosmic-rays interacting with molecular clouds.
In any case, the detection of Fe\emissiontype{I}~K$\alpha$ line is marginal, thus we need more detailed observations in the future.

\subsection{The abundances and mass of the Recombining Plasma}

The abundances and emission measure of RP given in table 3 are based on the H dominant plasma. We assume that the RP originates from an ejecta of $\sim$15\Mo~progenitor (figure~\ref{fig:mass}), which indicates super-solar abundances even for the light elements such as He, O and Ne. The estimations of emission measure and abundances of heavy elements largely depend on the abundances of He, O and Ne; large abundances
of these light elements enhance the continuum (mainly bremsstrahlung) flux, although the relative spectral shape is not affected because emission lines of He--Ne are absent in the relevant energy band of 1.2--7.0~keV.
We thus re-estimate the emission measure and abundances of heavy element assuming the light element mass distribution is the same as the progenitor star of 15\Mo. 
The abundances of Mg--Ca relative to Si in the RP component do not change from the original H dominant plasma but the averaged abundance of Si--Fe relative to hydrogen becomes $\sim$15 solar, confirming its ejecta origin. The emission measure becomes $n_{\rm e} n_{\rm H} V f \sim 2.1^{+0.2}_{-1.7}\times10^{58}~d_8^2$~cm$^{-3}$, where $f$ and $d_8$ are a filling factor and the SNR distance in the unit of 8~kpc, respectively. The ratio of the electron to the atomic hydrogen densities is  $n_{\rm e}/n_{\rm H} = 1.5$  and the number ratio of all the elements to hydrogen is 2.0.
Adopting the source radii of $7d_8$~pc, the ejecta mass is estimated to be $\sim 40~f^{1/2}~d_8^{5/2}$~\Mo. As is seen in Figure 1 of Chen \etal\ 2004, filling factor would be very small, hence we assume $f$ is in the range of 0.3--0.1.  Then the ejecta mass is $\sim$ 10--25~\Mo. This value is consistent with the initial assumption that 3C 391 is a remnant of CC SN with a progenitor mass of $\sim$ 15\Mo.

The ionization age of the RP is estimated using $n_{\mathrm{e}} t$. The electron density $n_{\mathrm{e}}$ is $0.9f^{-1/2}d_8^{-1/2}$~cm$^{-3}$.  
Then the elapsed time $t$ is $6.5\times10^3f^{-1/2}$ ($n_{\mathrm{e}}/0.9~\mathrm{cm}^{-3}$) years, which place this SNR as a rather old SNR. 
If we assume a small filing factor, the age of $1.7 \times 10^4$ yr estimated by Chen \& Slane (2001) has a very similar value.

\subsection{Origin of the Recombining Plasma}

Several scenarios are proposed to explain the formation of RPs in SNRs.
One is the rapid electron cooling by the thermal conduction from cold matters around SNRs (Kawasaki \etal\ 2002) or the adiabatic rarefaction (Itoh and Masai 1989, Shimizu \etal\ 2012).  The other is the ionization by the suprathermal or non-thermal particles (Masai \etal\ 2002, Ohnishi \etal\ 2011) or the high energy photons (Kawasaki \etal\ 2002).
Therefore, the presence of RP in 3C 391 means that the plasma experienced the rapid electron cooling or extra-ionization. 

We first discuss the electron cooling scenario.
If the electron cooling is due to thermal conduction, a timescale of the thermal conduction is 
\begin{equation}
t_{\rm cond} \sim 1.5 \times 10^{5}~\left( \frac{n_e}{0.9~{\rm cm}^{-3}} \right) \left( \frac{l}{7~{\rm pc}} \right)^{2} \left( \frac{k T}{0.50~{\rm keV}} \right)^{-\frac{5}{2}} d_8~{\rm yr},
\end{equation}
where $l$ is the temperature gradient scale length (Kawasaki et al. 2002). This timescale is much longer than the estimated age of 3C 391, therefore the thermal conductivity is unlikely to produce the whole RP in 3C 391 of which the radius is about 7~pc.

Next scenario is rarefaction, which is the possible origin of the RP in W28 (Sawada \& Koyama 2012). If the progenitor of 3C 391 was a massive star with a strong stellar wind activity, there can be a dense circumstellar media (CSM) surrounding the ejecta. The blastwave breaking out of the shell can lead to rarefaction of the PI or near CIE plasma. 
Thereafter the electron temperature can drop rapidly due to adiabatic expansion but the ionization states can stay more or less the same. 
This scenario is supported by the fact that the recombination timescale is close to the putative age of 3C 391. 
While Reynolds and Moffett (1993) estimated that the shell ``break-out" occurred at about 1600~yr after the supernova, the blast wave breaks out of the CSM at only $\sim$100~yr (Itoh and Masai 1989, Shimizu \etal\ 2012).
Since no significant spectral differences except the possible presence of a 6.4~keV line in the NW and a slightly larger $N_{\rm H}$ in the SE, we argue that the rarefaction of ejecta plasma was roughly symmetrical, unlike W49B as suggested by some models (e.g. Miceli et al 2010). Our results show that the RP of 3C 391 was produced by the ``break-out" from the CSM and evolved independently from the interstellar environment.
For 3C 391, we apply a model of an early phase break-out (Shimizu et al. 2012), while Zhou et al. 2011 proposed a later phase break-out for W49B.  This difference is due to their different circumstellar environments; 3C 391 would have a denser and more compact CSM compared to that of W49B.
Although the extra-ionization process could be an alternative interpretation, it is still impossible to identify the origin of RP with the present data alone. We hence encourage deeper observation of 3C 391.

\section{Conclusions}

We observed the middle-aged SNR 3C 391 with Suzaku for 100 ks. 
We generated a background model for the GRXE and analyzed the X-ray spectrum of 3C 391.
Our conclusions are summarized below:

\begin{itemize}
\item The X-ray spectrum of 3C 391 is well described by a CIE+RP model. 3C 391 has the highest $n_{\mathrm{e}}t$ value among the RP SNRs ever discovered. This means the RP of 3C 391 approaches ionization equilibrium the most.
\item Estimated progenitor mass of $\sim$15~\Mo~roughly agrees with the X-ray emitting mass of the RP derived from the corrected emission measure. This means only the ejecta plasma is overionized. 
\item The parameters of RP have no significant difference between the NW and SE regions. Thermal conduction is unlikely to be the origin of the RP at 3C 391. 
The symmetrical evolution of RP is consistent with a rarefaction of the plasma in the early phase.
In order to draw a more conclusive picture we need deeper observations with more sensitive instruments, such as those to be onboard the ASTRO-H mission (Takahashi \etal\ 2010).
\item Our detailed analysis of the spectra of 3C 391 suggests a hint of Fe\emissiontype{I}~K$\alpha$ emission at 6.4 keV ($\sim 2.4\sigma$), which can be explained by the interaction with a molecular cloud.
\end{itemize}

\bigskip

The authors express  sincere thanks  to all the Suzaku team members.
This work is supported by JSPS Scientific Research grant number 24540229 (K.K.). et al.


\begin{thebibliography}{00}
\bibitem[Abdo et al.(2010a)]{2010ApJ...712..459A} 
Abdo,~A.~A., et al. 2010a, \apj, 712, 459 

\bibitem[Abdo et al.(2010b)]{2010ApJ...718..348A}
Abdo,~A.~A., \etal\ 2010b, \apj, 718, 348 

\bibitem[Abdo et al.(2010c)]{2010ApJ...722.1303A}
Abdo,~A.~A., \etal\ 2010c, \apj, 722, 1303 

\bibitem[Abdo et al.(2010d)]{2010Sci...327.1103A}
Abdo,~A.~A., \etal\ 2010d, Science, 327, 1103 

\bibitem[Aharonian et al.(2004)]{2004Natur.432...75A} Aharonian,~F.~A., \etal\ 2004, \nat, 432, 75 

\bibitem[Anders 
\& Grevesse(1989)]{1989GeCoA..53..197A} Anders,~E., \& Grevesse,~N.\ 1989, \gca, 53, 197 

\bibitem[Arnaud(1996)]{1996ASPC..101...17A} Arnaud,~K.~A.\ 1996, 
Astronomical Data Analysis Software and Systems V, 101, 17

\bibitem[Brickhouse et al.(2000)]{2000ApJ...530..387B} Brickhouse,~N.~S., 
Dupree,~A.~K., Edgar,~R.~J., Liedahl,~D.~A., Drake,~S.~A., White,~N.~E., \& Singh,~K.~P.\ 2000, \apj, 530, 387 

\bibitem[Castro 
\& Slane(2010)]{2010ApJ...717..372C} Castro,~D., \& Slane,~P.\ 2010, \apj, 717, 372 

\bibitem[Chen 
\& Slane(2001)]{2001ApJ...563..202C} Chen,~Y., \& Slane,~P.~O.\ 2001, \apj, 563, 202 

\bibitem[Chen et al.(2004)]{2004ApJ...616..885C} Chen,~Y., Su,~Y., Slane,~P.~O., 
\& Wang,~Q.~D.\ 2004, \apj, 616, 885 

\bibitem[Condon et al.(1998)]{1998AJ....115.1693C} Condon,~J.~J., Cotton,~W.~D., 
Greisen,~E.~W., Yin,~Q.~F., Perley,~R.~A., Taylor,~G.~B., Broderick,~J.~J.\ 1998, \aj, 115, 1693 

\bibitem[Frail et al.(1996)]{1996AJ....111.1651F} Frail,~D.~A., Goss,~W.~M., 
Reynoso,~E.~M., Giacani,~E.~B., Green,~A.~J., \& Otrupcek,~R.\ 1996, \aj, 111, 1651 

\bibitem[Ishisaki et al.(2007)]{2007PASJ...59S.113I} Ishisaki,~Y., \etal\ 2007, \pasj, 59, 113 

\bibitem[Itoh \& Masai(1989)]{1989MNRAS.236..885I} Itoh,~H., \& Masai,~K.\ 1989, \mnras, 236, 885

\bibitem[Iwamoto et al.(1999)]{1999ApJS..125..439I} Iwamoto,~K., Brachwitz,~F., Nomoto,~K., Kishimoto,~N., Umeda,~H., Hix,~W.~R., \& Thielemann,~F.,\ 1999, \apjs, 125, 439 

\bibitem[Kaastra et al.(1996)]{1996uxsa.conf..411K} Kaastra,~J.~S., Mewe,~R., 
\& Nieuwenhuijzen,~H.\ 1996, UV and X-ray Spectroscopy of Astrophysical and Laboratory Plasmas, 411 

\bibitem[Kawasaki et al.(2002)]{2002ApJ...572..897K} Kawasaki,~M.~T., 
Ozaki,~M., Nagase,~F., Masai,~K., Ishida,~M., \& Petre,~R.\ 2002, \apj, 572, 897 

\bibitem[Kinugasa 
\& Tsunemi(1999)]{1999PASJ...51..239K} Kinugasa,~K., \& Tsunemi,~H.\ 1999, \pasj, 51, 239 

\bibitem[Koyama et al.(2007)]{2007PASJ...59S..23K} Koyama,~K., \etal\ 2007, \pasj, 59, 23

\bibitem[Kushino et al.(2002)]{2002PASJ...54..327K} Kushino,~A., Ishisaki,~Y., 
Morita,~U., Yamasaki,~N.~Y., Ishida,~M., Ohashi,~T., \& Ueda,~Y.\ 2002, \pasj, 54, 327 

\bibitem[Masai et al.(2002)]{2002ApJ...581.1071M} Masai, K., Dogiel, V.~A., 
Inoue, H., Sch{\"o}nfelder, V., \& Strong, A.~W.\ 2002, \apj, 581, 1071 

\bibitem[Miceli et 
al.(2010)]{2010A&A...514L...2M} Miceli,~M., Bocchino,~F., Decourchelle,~A., Ballet,~J., \& Reale,~F.\ 2010, \aap, 514, L2 

\bibitem[Mitsuda et al.(2007)]{2007PASJ...59S...1M} Mitsuda,~K., \etal\ 2007, \pasj, 59, 1

\bibitem[Morrison 
\& McCammon(1983)]{1983ApJ...270..119M} Morrison,~R., \& McCammon,~D.\ 1983, \apj, 270, 119 

\bibitem[Ohnishi et al.(2011)]{2011PASJ...63..527O} Ohnishi,~T., Koyama,~K., 
Tsuru,~T.~G., Masai,~K., Yamaguchi,~Hiroya., \& Ozawa,~M.,\ 2011, \pasj, 63, 527 

\bibitem[Ozawa et al.(2009)]{2009ApJ...706L..71O} Ozawa,~M., Koyama,~K., 
Yamaguchi,~H., Masai,~K., \& Tamagawa,~T.\ 2009, \apjl, 706, L71 

\bibitem[Radhakrishnan et al.(1972)]{1972ApJS...24...49R} Radhakrishnan,~V., 
Goss,~W.~M., Murray,~J.~D., \& Brooks,~J.~W.\ 1972, \apjs, 24, 49 

\bibitem[Reynolds 
\& Moffett(1993)]{1993AJ....105.2226R} Reynolds,~S.~P., \& Moffett,~D.~A.\ 1993, \aj, 105, 2226 

\bibitem[Sawada 
\& Koyama(2012)]{2012PASJ...64...81S} 
Sawada,~M., \& Koyama,~K.\ 2012, \pasj, 64, 81 

\bibitem[Serlemitsos et al.(2007)]{2007PASJ...59S...9S} Serlemitsos,~P.~J., \etal\ 2007, \pasj, 59, 9

\bibitem[Shimizu et al.(2012)]{2012PASJ...64...24S} Shimizu,~T., Masai,~K., 
\& Koyama,~K.\ 2012, \pasj, 64, 24 

\bibitem[Smith et al.(2001)]{2001ApJ...556L..91S} Smith,~R.~K., Brickhouse,~N.~S., 
Liedahl,~D.~A., \& Raymond,~J.~C.\ 2001, \apjl, 556, L91 

\bibitem[Takahashi et al.(2010)]{2010SPIE.7732E..27T} Takahashi,~T., \etal\ 2010, \procspie, 7732,  

\bibitem[Tawa et al.(2008)]{2008PASJ...60S..11T} Tawa,~N., \etal\ 2008, \pasj, 60, 11

\bibitem[Uchida et al.(2012)]{2012PASJ...64..141U} Uchida,~H., \etal\ 2012, \pasj, 64, 141 

\bibitem[Uchiyama et al.(2011)]{2011PASJ...63S.903U} Uchiyama,~H., 
Nobukawa,~M., Tsuru,~T., Koyama,~K., \& Matsumoto,~H.\ 2011, \pasj, 63, 903 

\bibitem[Uchiyama et al.(2013)]{2013PASJ...65...19U} Uchiyama,~H., 
Nobukawa,~M., Tsuru,~T.~G., \& Koyama,~K.\ 2013, \pasj, 65, 19 

\bibitem[Wilner et al.(1998)]{1998AJ....115..247W} Wilner,~D.~J., Reynolds,~S.~P., 
\& Moffett,~D.~A.\ 1998, \aj, 115, 247 

\bibitem[Woosley 
\& Weaver(1995)]{1995ApJS..101..181W} Woosley,~S.~E., \& Weaver,~T.~A.\ 1995, \apjs, 101, 181 

\bibitem[Yamaguchi et al.(2009)]{2009ApJ...705L...6Y} Yamaguchi,~H., Ozawa,~M., 
Koyama,~K., Masai,~K., Hiraga,~J.~S., Ozaki,~M., \& Yonetoku,~D.\ 2009, \apjl, 705, L6 

\bibitem[Yamaguchi et al.(2009)]{2009PASJ...61S.175Y} Yamaguchi,~H., Bamba,~A., 
\& Koyama,~K.\ 2009, \pasj, 61, 175

\bibitem[Yamaguchi et al.(2012)]{2012AdSpR..49..451Y} Yamaguchi,~H., Ozawa,~M.,
 \& Ohnishi,~T.\ 2012, Advances in Space Research, 49, 451

\bibitem[Yamauchi et al.(2013)]{2013PASJ...65....6Y} Yamauchi,~S., 
Nobukawa,~M., Koyama,~K., \& Yonemori,~M.\ 2013, \pasj, 65, 6 

\bibitem[Yasumi et al.(2014)]{2014arXiv1403.6898Y} Yasumi,~M., Nobukawa,~M., 
Nakashima,~S., Uchida,~H., Sugawara,~R., Tsuru,~T.~G., Tanaka,~T., Koyama,~K.\ 2014, \pasj, 59 

\bibitem[Zhou et al.(2011)]{2011MNRAS.415..244Z} Zhou,~X., Miceli,~M., 
Bocchino,~F., Orlando,~S., \& Chen,~Y.\ 2011, \mnras, 415, 244 

\end{thebibliography}

\end{document}